\documentclass[prd,onecolumn,aps,showpacs,nofootinbib,nobibnotes,superscriptaddress,preprintnumbers,notitlepage]{revtex4-1}

\usepackage[top=2.8cm, bottom=2.8cm, left=2.4cm, right=2.4cm]{geometry}
\usepackage[utf8]{inputenc}  
\usepackage[T1]{fontenc}     
\usepackage{amsmath,amssymb,lmodern} 
\usepackage{sidecap}
\usepackage[caption=false]{subfig}
\usepackage{graphicx}
\usepackage{tabularx}
\usepackage{enumerate} 
\usepackage{hyperref}
\usepackage{color}
\usepackage[dvipsnames]{xcolor}
\usepackage[capitalise]{cleveref}
\usepackage{here}
\usepackage{multirow}

\newcommand{\mr}[1]{\mathrm{#1}}
\newcommand{\mL}[1]{\mathcal{#1}}
\def\Ve{V_{\rm eff}}

\makeatletter
\renewcommand\@make@capt@title[2]{%
  \@ifx@empty\float@link{\@firstofone}{\expandafter\href\expandafter{\float@link}}%
   {\textbf{#1}}\@caption@fignum@sep#2\quad
}%
\makeatother

\newcommand{\ba}{\begin{eqnarray}}
\newcommand{\ea}{\end{eqnarray}}
\newcommand{\be}{\begin{equation}}
\newcommand{\ee}{\end{equation}}
\newcommand{\bea}{\begin{eqnarray}}
\newcommand{\eea}{\end{eqnarray}}
\newcommand{\beq}{\begin{equation}}
\newcommand{\eeq}{\end{equation}}
\newcommand{\beqar}{\begin{eqnarray}}
\newcommand{\eeqar}{\end{eqnarray}}
\newcommand{\beqars}{\begin{eqnarray*}}
\newcommand{\eeqars}{\end{eqnarray*}}
\newcommand{\bc}{\begin{center}}
\newcommand{\ec}{\end{center}}
\newcommand{\ben}{\begin{enumerate}}
\newcommand{\een}{\end{enumerate}}
\newcommand{\bit}{\begin{itemize}}
\newcommand{\eit}{\end{itemize}}
\newcommand{\bw}{\begin{widetext}}
\newcommand{\ew}{\end{widetext}}
\newcommand{\bcl}{\begin{columns}}
\newcommand{\ecl}{\end{columns}}

\begin{document}
\begin{flushright}
	PI/UAN-2020-666FT
 \end{flushright}

\title{Topological mass generation and $2-$forms}

\author{Juan P.~Beltr\'an Almeida} \email{jubeltrana@unal.edu.co} 
\affiliation{Universidad Nacional de Colombia, Sede Bogot\'a, Facultad de Ciencias, Departamento de F\'isica,  Av. Cra 30 No 45-03, 
Bogot\'a DC, Colombia}
\affiliation{Departamento de F\'isica, Facultad de Ciencias, Universidad Antonio Nari\~no,  Cra 3 Este \# 47A-15, 
Bogot\'a DC, Colombia}
\author{Alejandro Guarnizo}\email{alejandro.guarnizo@correounivalle.edu.co} 
\affiliation{Departamento de F\'isica, Facultad de Ciencias, Universidad Antonio Nari\~no,  Cra 3 Este \# 47A-15, 
Bogot\'a DC, Colombia}
\affiliation{Departamento de F\'isica, Universidad del Valle,
Ciudad Universitaria Mel\'endez, Santiago de Cali 760032, Colombia}

\author{Lavinia Heisenberg}
\email{lavinia.heisenberg@phys.ethz.ch}
\affiliation{Institute for Theoretical Physics,
ETH Zurich, Wolfgang-Pauli-Strasse 27, 8093, Zurich, Switzerland}

\author{C\'esar A.~Valenzuela-Toledo} \email{cesar.valenzuela@correounivalle.edu.co}
\affiliation{Departamento de F\'isica, Universidad del Valle,
Ciudad Universitaria Mel\'endez, Santiago de Cali 760032, Colombia}

\author{Jann Zosso} \email{jzosso@phys.ethz.ch}
\affiliation{Institute for Theoretical Physics,
ETH Zurich, Wolfgang-Pauli-Strasse 27, 8093, Zurich, Switzerland}

\date{\today}

\begin{abstract}
In this work we revisit the topological mass generation of 2-forms and establish a connection to the unique derivative coupling arising in the quartic Lagrangian of the systematic construction of massive $2-$form interactions, relating in this way BF theories to Galileon-like theories of 2-forms. In terms of a massless $1-$form $A$ and a massless $2-$form $B$, the topological term manifests itself as 
the interaction $B\wedge F$, where $F = {\rm d} A$ is the field strength of the $1-$form.  Such an interaction leads to a mechanism of generation of mass, usually referred to as ``topological generation of mass'' in which the single degree of freedom propagated by the $2-$form is absorbed by the $1-$form, generating a massive mode for the $1-$form. Using the systematical construction in terms of the Levi-Civita tensor, it was shown that, apart from the quadratic and quartic Lagrangians, Galileon-like derivative self-interactions for the massive 2-form do not exist. A unique quartic Lagrangian $\epsilon^{\mu\nu\rho\sigma}\epsilon^{\alpha\beta\gamma}_{\;\;\;\;\;\;\sigma}\partial_{\mu}B_{\alpha\rho}\partial_{\nu}B_{\beta\gamma}$ arises in this construction in a way that it corresponds to a total derivative on its own but ceases to be so once an overall general function is introduced. We show that it exactly corresponds to the same interaction of topological mass generation. Based on the decoupling limit analysis of the interactions, we bring out supporting arguments for the uniqueness of such a topological mass term and absence of the Galileon-like interactions. Finally, we discuss some preliminary applications in cosmology.    
\end{abstract}

%\pacs{04.50.Kd, 04.70.Bw}

\maketitle

%%%%%%%%%%%%%%%%%%%%%%%%%%%%%%
\section{Introduction}
%%%%%%%%%%%%%%%%%%%%%%%%%%%%%
The successful construction of Galileon theories \cite{Nicolis:2008in} has changed our perspective on standard effective field theories (EFTs). Even though the resulting Lagrangian contains higher derivative terms, the equations of motion remain second order and hence avoid the Ostrogradski instability. For astrophysical applications these higher order operators generically need to be large. From a standard EFT point of view, this would be disastrous as it would rely on `irrelevant operators'' becoming large and thus going beyond the regime of validity of the theory. For the Galileon theories, however, this is different. The operators are rearranged in a way that higher order derivative operators with second order equations of motion can become relevant whereas operators with even more derivatives per field giving rise to higher order equations of motion are treated as irrelevant. Furthermore, this reorganization of the operators is stable under quantum corrections \cite{Luty:2003vm,Nicolis:2004qq,Heisenberg:2019udf,Heisenberg:2019wjv}. This is the non-renormalization theorem of the Galileon. For this, it is crucial that the Galileon symmetry is only realized up to total derivatives \cite{deRham:2012ew}.

A similar attempt to construct such Galileon-like Lagrangians for arbitrary p-forms immediately met a no-go theorem in four dimensions \cite{Deffayet:2010zh}. This includes massless 1-form. Hence, derivatives acting on the field strength of a Maxwell field do not permit the construction of Lagrangians with second order equations of motion and gauge invariance. However, this obstruction does not apply to the case of massive spin-1 fields. The removal of gauge invariance allows the construction of non-trivial Galileon-like derivative self-interactions of the massive vector field with three propagating degrees of freedom, the generalized Proca theories \cite{Heisenberg:2014rta,Allys:2015sht,Jimenez:2016isa} (see also \cite{Heisenberg:2018vsk}). They represent rich phenomenology in cosmological and astrophysical applications \cite{Tasinato:2014eka,DeFelice:2016yws,DeFelice:2016cri,DeFelice:2016uil,Heisenberg:2016wtr,deFelice:2017paw,Heisenberg:2017hwb,Chagoya:2017fyl}. Concerning Galileon-like Lagrangians for a Kalb-Rammond field, a massless 2-form, it is only possible to construct such derivative self-interactions starting from seven dimensions. Even if the gauge symmetry is removed, the difficulty persists. It was shown in \cite{Heisenberg:2019akx}, that only interactions belonging to the quadratic and quartic Lagrangians can successfully be constructed. Within the quartic Lagrangians, a unique interaction $\epsilon^{\mu\nu\rho\sigma}\epsilon^{\alpha\beta\gamma}_{\;\;\;\;\;\;\sigma}\partial_{\mu}B_{\alpha\rho}\partial_{\nu}B_{\beta\gamma}$ manifests itself as a total derivative, which becomes non-trivial with an overall general function of the 2-form norm. We show that this special interaction corresponds to the interaction $B\wedge F$ of BF theories \cite{Blau:1989dh,Blau:1989bq,Horowitz:1989ng} which gives rise to ``topological mass generation''  \cite{Allen:1990gb,Quevedo:1996uu,Dvali:2005ws}. This ``topological mass generation'' refers to the fact that the single degree of freedom propagated by a massless $2-$form is absorbed by a $1-$form, generating a massive mode for the $1-$form. This mechanism was revisited in  \cite{Almeida:2018fwe,Almeida:2019xzt} when looking for particular models of coupled $p-$forms suitable for cosmological applications such as inflation and dark energy. \\

This paper is organized as follows. In section \ref{sec:syspf} we revisit the results from \cite{Almeida:2018fwe,Almeida:2019xzt} about interacting $p-$forms paying special attention to the interaction between the $1$ and $2$-forms through the topological term $B\wedge F$. We show its direct relation to the unique Galileon-like term of the systematic construction of massive $2-$form interactions. This allows us to build a direct duality between BF theories and Galileon-like theories. In section \ref{g_construction} we recall the systematic Galileon-like construction carried out in \cite{Heisenberg:2019akx} where a new kind of interaction for the massive $2-$form is found. Then, in section \ref{decoupling} we discuss the decoupling limit of the system, that gives supporting arguments for the uniqueness of such a topological mass term and absence of the Galileon-like interactions. Finally, in section \ref{capp} we show some simple cosmological applications aiming to emphasize the relevance and to boost the interest for $2-$forms models  applied in cosmological setups. 

%%%%%%%%%%%%%%%%%%%%%%%%%%%%%%%%%%%
\section{ $B\wedge F$ term and the topological generation of mass  }
\label{sec:syspf}
%%%%%%%%%%%%%%%%%%%%%%%%%%%%%%%%%%%

In this section we briefly recall the motivations and results from \cite{Almeida:2018fwe,Almeida:2019xzt}. In that reference, the authors discussed general models of interacting $p-$form Lagrangians subject to the following restrictions: 1) $U(1)$ gauge invariance, 2) up to first order derivatives of the $p-$forms in the Lagrangian, 3) up to cubic terms in the derivatives of the $p-$forms, 4) having a Hamiltonian bounded from below, 5) hyperbolicity of the equations of motion \cite{Fleury:2014qfa}.  With the restrictions mentioned before, it was found that, in four dimensions, the more general action involving general interactions between the $p-$forms is: 
\begin{equation}
\mL{S} = \int \mr{d}^4 x \sqrt{-g}  \left[  \frac{M_{\mr{pl}}^2}{2}R   -\frac{1}{2}\partial_{\mu}\phi \partial^{\mu}\phi  - \Ve(\phi)   - \frac{1}{4} f_{1}(\phi) F^2 -  \frac{1}{12} f_{2}(\phi) H^2 -  \frac{1}{4} g_{1}(\phi) F \tilde{F} - \frac{1}{2} m  B\tilde{F} \right]\,,
\label{eq:LT}
\end{equation}
where the $1-$ and $2-$forms are $A_{\mu}$ and $B_{\mu \nu}$ respectively, and their  field strengths are 
\begin{equation}
 F_{\mu  \nu}   \equiv  2 \partial_{[\mu } A_{\nu]}\,, \qquad 
H_{\mu \nu \alpha} \equiv 3  \partial_{[\mu} B_{\nu \alpha]} \,.
\end{equation} 
Additionally, we used the shorthand notation 
\begin{equation}
F^2 \equiv F_{\mu  \nu}F^{ \mu  \nu}\,,\quad
H^2  \equiv H_{\mu \nu \alpha}H^{\mu \nu \alpha}\,,\quad   F \tilde{F} \equiv F_{\,\mu  \nu}\tilde{F}^{\mu  \nu}  \, ,\quad
B\wedge F = B \tilde{F} \equiv B_{\,\mu  \nu }\tilde{F}^{\mu  \nu}\,.  
\end{equation}
In the previous action, the functions $f_{i}(\phi)$ are arbitrary functions of the fields, only restricted to be positive definite $f_i >0$, and $m$ is constant in order to preserve the gauge invariance. The coupling function $g_1(\phi)$ is not restricted by any of the conditions mentioned before. The effective potential $\Ve$ is induced by the coupling of the $3-$form field and the scalar field \cite{Kaloper:2008qs,Kaloper:2008fb, Bielleman:2015ina,Ibanez:2015fcv,Valenzuela:2016yny,Farakos:2017jme}.  The term $B \wedge{F}$ \cite{Blau:1989dh,Blau:1989bq,Horowitz:1989ng} is responsible for the mechanism of topological generation of mass described in detail in \cite{Allen:1990gb,Quevedo:1996uu,Dvali:2005ws}.  

\subsection{Relating BF with Galileon-like interactions}\label{sec_relBFgal}
Let's focus now in a model involving only $1$ and $2-$forms without introducing an extra scalar field degree of freedom, and let's restrict our analysis to flat space (some discussion about curved space and non minimal coupling with gravity can be found in \cite{Heisenberg:2019akx}. For complementary discussion on this topic see also \cite{Yoshida:2019dxu,Takahashi:2019vax,Koivisto:2009sd}). As found in \cite{Almeida:2018fwe}, the only coupling term involving $1$ and $2-$forms consistent with the five restrictions mentioned before, particularly with the condition of being gauge invariant for $A_{\mu}$ and $B_{\mu \nu}$ is the term $B_{\,\mu  \nu }\tilde{F}^{\mu  \nu}$. With this, the model we will consider is written with the Lagrangian  
 %Here we focus on the analysis of the $2-$forms sector of the Lagrangian \eqref{eq:LT}. To this end, we identify the scalar field in \eqref{eq:LT} with the modulus squared of the $2-$form as $\phi = B^2 \equiv B_{\mu\nu}B^{\mu\nu}$ and we  use the equation of motion of the $2-$form derived from \eqref{eq:LT} 
\begin{equation}
\mL{S}_{AB} = -\int \mr{d}^4 x \left[       \frac{1}{4}   F^2 +  \frac{1}{12}   H^2  + \frac{1}{2} m  B\tilde{F} \right]\,,
\label{ABm}
\end{equation} 
where $m$ is a constant in order to preserve gauge invariance. From this, we derive the equation of motion for the $2-$form
\begin{equation} 
\label{A2eq}
\partial^{\mu} H_{\mu\nu \alpha}   -m  \tilde{F}_{\nu\alpha}  =0 \,,
\end{equation}
%The previous equation is complemented with the Bianchi identities:
%\begin{equation}
%\nabla_{\mu} \tilde{F}^{\mu\nu} = 0,\quad \mbox{and} \quad \nabla_{\mu} \tilde{H}^{\mu} = 0.
%\end{equation}
which can be formally solved for $\tilde{F}^{\mu\nu}$ and $F^{\mu\nu}$ obtaining
\begin{equation}\label{solft}
\tilde{F}_{\mu\nu} = \frac{1}{m } \partial^{\alpha} H_{\alpha\mu \nu}\,,  \quad F^{\mu\nu} = -\frac{1}{6 m }\partial^{[\mu}   \tilde{H}^{\nu ]}\,. 
\end{equation} 
%\begin{equation}\label{solf}
%F^{\mu\nu} = -\frac{1}{3m }\nabla^{[\mu} f_2(B^2) \tilde{H}^{\nu ]}. 
%\end{equation} 
Using this formal solution we can rewrite the Lagrangian \eqref{ABm} only in terms of the $2-$form. 
%By doing that, we see that the $F^2$ can be expressed as 
%\be
%F^2 \propto \nabla^{[\mu} f_2(\phi) \tilde{H}^{\nu ]} \nabla_{[\mu} f_2(\phi) \tilde{H}_{\nu ]}.
%\ee
%This term can be expressed as a combinations of $\partial_\mu \phi \partial^{\mu}\phi$ and $H^2$. On the other hand, 
The structure of the interacting term $B\wedge F$ is of particular interest. Using \eqref{solft} we can write the term $B\wedge F$ as follows
\begin{equation}
B\wedge F =   B^{\mu\nu} \tilde{F}_{\mu\nu}    =   \frac{1}{m }  B_{\mu\nu}  \partial^{\alpha}\left( H_{\alpha\mu \nu}  \right) =  \frac{ 1 }{ m }  B^{\mu\nu} \left[  \partial^{\alpha} \partial_{\alpha} B_{\mu\nu}  +   \partial^{\alpha} \partial_{\mu} B_{\nu \alpha} + \partial_{\nu }  \partial^{\alpha}B_{ \alpha \mu}  \right] \,,
\end{equation}
where we exchanged the order of the derivatives in the last term. Now, integrating by parts the terms with second derivatives of the $2-$form we obtain 
\begin{equation}
B\wedge F = - \frac{  1 }{m }    \left[  \partial^{\alpha}   B^{\mu\nu}   \partial_{\alpha} B_{\mu\nu}  +  \partial^{\alpha}  B^{\mu\nu}  \partial_{\mu} B_{\nu \alpha} + \partial_{\nu}  B^{\mu\nu}  \partial^{\alpha}   B_{\alpha \mu} \right]\,, 
\end{equation}
which can be reordered in the form
\begin{equation}
B\wedge F =  - \frac{  1 }{m }    \left[  \partial^{\alpha}   B^{\mu\nu}   \partial_{\alpha} B_{\mu\nu}  +  \partial^{\alpha}  B^{\mu\nu}  \partial_{\mu} B_{\nu \alpha} - \partial^{\alpha}  B^{\mu\nu}  \partial_{\nu} B_{\mu \alpha} \right] - \frac{  1 }{m }    \left[   \partial^{\alpha}  B^{\nu\mu}  \partial_{\mu} B_{\nu \alpha} + \partial_{\nu}  B^{\nu\mu}  \partial^{\alpha}   B_{ \mu \alpha } \right]\,.  
\end{equation}  
We can identify the term inside the first brackets as $H^2/3$, so, it is absorbed in the kinetic Maxwell like term for the $2-$form. On the other hand, the term in the second brackets can be recognized as the novel interaction term 
\begin{equation}
{\cal L}_4^{T} =  \partial^{\alpha}  B^{\nu\mu}  \partial_{\mu} B_{\nu \alpha} + \partial_{\nu}  B^{\nu\mu}  \partial^{\alpha}   B_{ \mu \alpha } 
\end{equation} 
found in the systematic construction carried out in reference \cite{Heisenberg:2019akx} (see  \cref{modKinL4} in next section).  To summarize, we can write the topological term $ B\wedge F$ as follows 
\begin{equation}
B\wedge F  = - \frac{  1 }{m }   {\cal L}_4^T - \frac{  1 }{3 m }  H^{\alpha \mu \nu} H_{\alpha \mu\nu}\,.   
\end{equation}  
As warned in \cite{Heisenberg:2019akx}, the term ${\cal L}_4^{T}$ is a total derivative as can be checked after integration by parts twice. Nevertheless,  ${\cal L}_4^{T}$ is not a total derivative anymore when it is multiplied by an arbitrary function $f_4(B^2)$ (see \cref{modKinL4} below) where  $B^2=B_{\mu\nu}B^{\mu\nu}$.    
The identification of ${\cal L}_4^{T}$ as part of the topological term $B\wedge F$ provides a link between the construction of Galileon-like derivative self interactions developed in \cite{Heisenberg:2019akx}  and briefly recalled in \ref{g_construction}, and BF theories (specially the approach followed in \cite{Almeida:2018fwe, Almeida:2019xzt}).  
%%%%%%%%%%%%%%%%%%%%%%%%%%%%%%%%%%%
\subsection{ Including the duals in the systematic construction }
\label{topological}
%%%%%%%%%%%%%%%%%%%%%%%%%%%%%%%%%%%
%Here we list all the topological terms which are present in a theory with $1-$ and $2-$forms fields. 
%\be
 %{\cal S}=  \int   \left[  \alpha_1 B\wedge B +    \alpha_2 B\wedge F  + \alpha_3 F\wedge F \right]\,,
%\ee
If we allow the possibility of including parity breaking terms, we could also consider the inclusion of the duals of the $2-$form and its field strength, this is  $*B$ and $*H$, in the systematic construction. We follow here closely the discussion of the example of ``compact QED'' presented in \cite{Quevedo:1996uu}.  Beside  the topological term $B\wedge F$ that we discussed before,  the possible non vanishing contributions that can be constructed with those objects are
\begin{equation}\label{LB}
  \mL{L}_{B}=  \int   \left[  a_1 H\wedge *H +    a_2  B \wedge *B  + a_3 B\wedge B \right] =   \int  \mr{d}^4 x      \left[  a_1 H_{\mu\nu\sigma}   H^{\mu\nu\sigma}  +    a_2  B_{\mu\nu }   B^{\mu\nu}   + a_3  B_{\mu\nu }   \tilde{B}^{\mu\nu}  \right]\,,
\end{equation}
where $\tilde{B}^{\mu\nu} = \epsilon^{ \mu\nu\sigma\rho} B_{\sigma\rho }/2  $ and we consider constant coefficients $a_1, a_2, a_3$. %Other possible contractions vanish, such as $H_{\mu\nu\sigma} B^{\mu\nu} \tilde{H}^{\sigma}$ or are equivalent to the previous three combinations such as $\tilde{H}^{\mu}\tilde{H}_{\mu}$. 
The equations of motion derived from this action are
\begin{equation}\label{eom2form}
6 a_1 \partial_{\mu} H^{\mu\nu\sigma}  - 2 a_2  B^{ \nu \sigma}  -  2 a_3 \frac{  \epsilon^{  \nu\sigma \mu \rho}  }{2   }  B_{ \mu \rho} = 0\,.
\end{equation}
We can derive the previous expression with respect to $ \partial_{\nu} $ and obtain 
\begin{equation}
a_2  \partial_{\nu} B^{ \nu \sigma}  +  a_3 \frac{  \epsilon^{  \nu\sigma \mu \rho}  }{2     } \partial_{\nu} B_{ \mu \rho} =  0\,,
\end{equation}
which can be arranged in the form  
\begin{equation}\label{BTH}
    a_2  \partial_{\nu} B^{ \nu \sigma}  -  a_3 \tilde{H}^{\sigma}  = 0,\quad \mbox{with} \quad \tilde{H}^{\sigma} \equiv  \frac{  \epsilon^{ \sigma \nu \mu \rho}  }{6    } \partial_{[\nu} B_{ \mu \rho]}= \frac{  \epsilon^{ \sigma \nu \mu \rho}  }{6     } H_{\nu \mu \rho}\,.
\end{equation}
We further apply the exterior derivative ${\mr{d} }\wedge = \epsilon_{\alpha\nu\sigma\rho} \partial^{\alpha}$ to the equation of motion \cref{eom2form} and use \cref{BTH} to obtain
\begin{equation}
12 a_1\left(  \partial_{\beta} \partial^{\mu} \tilde{H}_{\mu}   - \partial_{\mu} \partial^{\mu} \tilde{H}_{\beta}  \right)   + 4 a_2  \tilde{H}_{\beta}   +  4 \frac{a_3^2}{a_2}  \tilde{H}_{\beta} = 0\,.
\end{equation}
The term $ \partial^{\mu} \tilde{H}_{\mu}$ vanishes due to the Bianchi identity and we are left with
\begin{equation}\label{masstH}
   \left( \partial_{\mu} \partial^{\mu}  + m^2  \right)   \tilde{H}_{\beta} = 0\,, \quad \mbox{with} \quad m^2 =   -  \frac{ a_2}{3 a_1} \left( \frac{a_3^2}{ a_2^2}   + 1\right)\,,    
\end{equation}
which we  recognize as the equation of motion for a massive vector field $\tilde{H}_{\beta}$.  In this sense, the particle content of the Lagrangian \cref{LB} is the same as the content of a model with a massive vector field $*H$. Despite the fact that this vector is defined as an axial vector, and the topological mass term $ B\wedge B$ seems to be a source of parity breaking, the theory remains parity conserving. Then, the presence of the term $B\wedge B$ provides another source of a  mass term into the theory just like the topological $B\wedge F$ term does. This situation would change if we consider general non linear functions $F(U, V, W)$ where $U=H_{\mu\nu\sigma}   H^{\mu\nu\sigma}$, $V= B_{\mu\nu }   B^{\mu\nu} $ and $W=B_{\mu\nu }   \tilde{B}^{\mu\nu}$. In such case, the equation of motion and the Bianchi identity leads to 
\begin{align}\label{eq:nonlineartB}
& 3 \left[ \partial_{\beta} \partial^{\gamma} \left( F_{U} \tilde{H}_{\gamma} \right)-  \partial^{\gamma} \partial_{\gamma}  \left( F_{U} \tilde{H}_{\beta} \right) \right] 
+ \left( F_{V} + \frac{F_W^2}{F_{V}}  \right) \tilde{H}_{\beta} \nonumber \\ 
&  -  \left( \partial^{\gamma}F_{V} + \frac{F_W}{F_{V}}  \partial^{\gamma}F_{W} \right) \tilde{B}_{\gamma \beta}  +  \left( \partial^{\gamma}F_{W} - \frac{F_W}{F_{V}}  \partial^{\gamma}F_{V} \right) {B}_{\gamma \beta} =0\,,
\end{align}
where $F_{U} = \partial_{U} F$, etc. As can be checked, the previous equation reduces to \cref{masstH} when the function $F$ is linear on their arguments, and then, leads to a parity conserving system. For non linear $F$ we have a parity breaking situation as evidenced from the presence of the third and fourth terms in \cref{eq:nonlineartB}. Nevertheless, as we will not consider parity breaking models, in the following  we will neglect non linear Lagrangians in the argument $W$.  \\
In the next section, we will revisit the Galileon-like construction for massive $2-$forms performed in \cite{Heisenberg:2019akx}, to which we had linked the coupled 1 and $2-$form discussed in the subsection \ref{sec_relBFgal}. 
 
% With this we complete our discussion about the connection between the approaches followed in \cite{Heisenberg:2019akx} and \cite{Almeida:2018fwe}. 
%%%%%%%%%%%%%%%%%%%%%%%%%%%%%%%%%%%%%%%%%%%%%%

%%%%%%%%%%%%%%%%%%%%%%%%%%%%%%%%%%%%%%%%%%%%%%
\section{Systematical construction}
\label{g_construction}
%%%%%%%%%%%%%%%%%%%%%%%%%%%%%%%%%%%%%%%%%%%%%%
   The fact that one cannot construct derivative Galileon interactions beyond the trivial cases can be shown in two complementary and independent ways. The first one is using the systematical construction in terms of the Levi-Civita tensor and the second one is the decoupling limit analysis. This was already studied in detail in \cite{Heisenberg:2019akx}. We summarize the argument based on the systematical construction in this section and give more detail on the decoupling limit analysis in the next section. 
   In \cite{Heisenberg:2019akx} the antisymmetry properties of the Levi-Civita tensor was used to investigate the possible construction of Galileon-type interactions for the massive 2-form with 3 propagating degrees of freedom. It was found, that besides the $\mathcal{L}_2$ and $\mathcal{L}_4$ Lagrangians, it is not possible to construct derivative self-interactions for the massive 2-form in four dimensions. The quadratic Lagrangian is simply a combination of potential-like and gauge-invariant quantities
   \begin{equation} \label{LagL2}
\mathcal{L}_{2} = f_{2}(B_{\mu\nu}, H_{\mu\nu\rho}, \bar{H}_{\mu})\,.
\end{equation}
The higher order Lagrangians are constructed systematically in terms of powers of the fundamental object $\partial_{\alpha}B_{\mu\nu}$ together with two Levi-Civita tensors $ f(B^2)\epsilon\epsilon(\partial B)^{m}B^{n}$. Since $\partial B$ carries 3 but the Levi-Civita tensor 4 indices, the Lagrangian $\mathcal{L}_{3}$ cannot be constructed
\begin{equation} \label{LagL3}
\mathcal{L}_{3} = 0\,.
\end{equation}
This problem of even versus an odd number of indices is avoided in the quartic Lagrangian $\mathcal{L}_{4}$. Ignoring contributions that belong to $\mathcal{L}_2$ (like the kinetic term), the only non-trivial term that we can construct at this order is
\begin{align}\label{LagL4}
\mathcal{L}_{4}^T&=\epsilon^{\mu\nu\rho\sigma}\epsilon^{\alpha\beta\gamma}_{\;\;\;\;\;\;\sigma}\partial_{\mu}B_{\alpha\rho}\partial_{\nu}B_{\beta\gamma} \, , \nonumber\\
&=\partial_\mu B^{\mu\nu}\partial_\alpha B_\nu{}^{\alpha}+\partial_\nu B_{\mu\alpha}\partial^\alpha B^{\mu\nu}\,.
\end{align}
Without multiplying this contraction with an overall function $f_4(B^2)$, it corresponds to a total derivative and can be directly related to $B\wedge F$, as shown in the section \ref{sec_relBFgal}.
Hence, at this order this is the only non-trivial genuinely new term, that cannot be absorbed into $\mathcal{L}_2$
\begin{equation}\label{modKinL4}
\mathcal{L}_{4}^{0B}=f_4(B^2)\Big(\partial_\mu B^{\mu\nu}\partial_\alpha B_\nu{}^{\alpha}+\partial_\nu B_{\mu\alpha}\partial^\alpha B^{\mu\nu} \Big)\,.
\end{equation}
This term is quite special. It looks like a modified kinetic term without gauge invariance but as we saw above can be directly related to the topological mass term. We could also construct contractions higher in $n$. For instance,
\begin{align} 
\mathcal{L}_{4}^{(1B)}&=
 \epsilon^{\mu\nu\rho\sigma}\epsilon^{\alpha\beta\gamma\delta}\partial_{\mu}B_{\alpha\rho}\partial_{\nu}B_{\beta\gamma} B_{\sigma\delta}\,, \\
 \mathcal{L}_{4}^{(2B)}&=
 \epsilon^{\mu\nu\rho\sigma}\epsilon^{\alpha\beta\gamma\delta}\partial_{\mu}B_{\alpha\rho}\partial_{\nu}B_{\beta\gamma} B_{\sigma\lambda}B^\lambda{}_\delta \, , 
\end{align} 
multiplied by a general function of the 2-form norm, respectively. Going to higher order contributions in $(\partial B)$ is not possible beyond this order since the two Levi-Civita tensors
contain eight indices and $(\partial B)^3$ would require 9 and $(\partial B)^4$ 12 indices and so on. Therefore we have
\begin{equation}
\mathcal{L}_{i}=0 \qquad \text{for} \qquad i\geqq5 \, .
\end{equation}
Hence, the systematical construction stops at $\mathcal{ L}_{4}$ and we cannot construct Galileon interactions for the massive 2-form beyond $\mL{L}_{2}$ and $\mL{L}_{4}$. 
Additional support for this difficulty of constructing Galileon-type derivative interactions also comes from the decoupling limit analysis, which we will discuss in the next section.

%%%%%%%%%%%%%%%%%%%%%%%%%%%%%%%%%
 \section{Decoupling limit analysis}
 \label{decoupling}
%%%%%%%%%%%%%%%%%%%%%%%%%%%%%%%%%

  The decoupling limit analysis already reveals important conditions about the allowed interactions once the gauge symmetry is restored using the Stueckelberg trick. For this purpose, we perform the following change of variables
  \begin{equation}\label{TrafoStuck}
  B_{\mu\nu} \to B_{\mu\nu}+\frac{1}{m}\partial_{[\mu}A_{\nu]}\,,
  \end{equation}
  where the massless spin-1 field $A_\mu$ represents the Stueckelberg field. The original massive 2-form propagates 3 propagating degrees of freedom. After reintroducing the Stueckelberg field, the massive 2-form decomposes into a massless 1-form and a massless 2-form, still propagating $1+2$ degrees of freedom. In the decoupling limit, where we sent the mass of the 2-form to zero, we obtain two decoupled massless 1- and 2-form. In order to illustrate that, we take the standard Lagrangian of a massive 2-form
    \begin{equation}
  \mathcal{L}=-\frac1{12} H_{\mu\nu\rho} H^{\mu\nu\rho} -\frac{m^2}4 B_{\mu\nu}B^{\mu\nu} \,,
  \end{equation}
  where $H$ is the field strength of the massive 2-form, and perform the transformation in \eqref{TrafoStuck} to it. The kinetic term is immun to it but the mass term changes into
      \begin{equation}
  \mathcal{L}=-\frac1{12} H_{\mu\nu\rho} H^{\mu\nu\rho} -\frac{m^2}4 \left(B_{\mu\nu}+\frac{1}{m}\partial_{[\mu}A_{\nu]}\right)^2 \,.
  \end{equation}
  In the limit $m\to0$, we obtain a massless 1-form and a massless 2-form, decoupled from each other and each of them is invariant under gauge symmetries
        \begin{equation}
  \mathcal{L}=-\frac1{12} H_{\mu\nu\rho} H^{\mu\nu\rho} -\frac14 F_{\mu\nu}F^{\mu\nu} \,,
  \end{equation}
  where $F$ is the field strength of the massless 1-form. If there are genuinely new Galileon interactions for the original massive 2-form, we would see their presence in the decoupling limit. We would need to construct Galileon interactions for the massless 2-form and massless 1-form sectors in the decoupling limit. Since the 1-form and the 2-form have gauge symmetries, they can manifest themselves only through the gauge invariant field strengths $F$ and $H$. For the massless 1-form we would be after interactions of the form
\begin{equation}\label{Galmassless1form}
  \mathcal{L}=\epsilon^{\mu_1\mu_2\cdots}\epsilon^{\nu_1\nu_2\cdots}F_{\mu_1\mu_2} F_{\nu_1\nu_2}\left( \partial_{\mu_k}F_{\nu_l\nu_{l+1}}\cdots \right)\left( \partial_{\nu_j}F_{\mu_m\mu_{m+1}}\cdots\right) \,.
  \end{equation}
  Since $F$ comes in with 2 indices, we can only start constructing such terms for dimensions $D\ge 5$. However, they do not correspond to any genuinely new interaction, since they contribute in the form of a total derivative. Hence, the interactions in \cref{Galmassless1form} can be rewritten as
      \begin{equation}
  \mathcal{L}=\frac12\partial_{\nu_j}\left\{\epsilon^{\mu_1\mu_2\cdots}\epsilon^{\nu_1\nu_2\cdots}F_{\mu_1\mu_2} F_{\nu_1\nu_2}F_{\nu_m\nu_{m+1}}\left( \partial_{\mu_k}F_{\nu_l\nu_{l+1}}\cdots \right)\left( \partial_{\nu_i}F_{\mu_n\mu_{n+1}}\cdots\right) \right\} \,.
  \end{equation}
  Therefore, this constitutes a no-go theorem for a massless 1-form to have Galileon-type derivative self-interactions in any dimensions \cite{Deffayet:2013tca}. Hence, there is no way to construct Galileon interactions for the massless 1-form in our decoupling limit. A similar no-go also exists for the massless 2-form, in four dimensions. Similarly, this time we are after following type of interactions for the massless 2-form
\begin{equation}\label{Galmassless2form}
  \mathcal{L}=\epsilon^{\mu_1\mu_2\mu_3\cdots}\epsilon^{\nu_1\nu_2\nu_3\cdots}H_{\mu_1\mu_2\mu_3} H_{\nu_1\nu_2\nu_3}\left( \partial_{\mu_k}H_{\nu_{l-1}\nu_l\nu_{l+1}}\cdots \right)\left( \partial_{\nu_j}H_{\mu_{m-1}\mu_m\mu_{m+1}}\cdots\right) \,.
  \end{equation}
  Since $H$ comes in with 3 indices, such construction of interactions is only possible for $D\ge7$ \cite{Deffayet:2010zh}. For instance, in seven dimensions we can construct $\mathcal{L}^{(D=7)}=\epsilon^{\mu\nu\rho\sigma\tau\phi\chi}\epsilon^{\alpha\beta\gamma\delta\epsilon\xi\eta}H_{\mu\nu\rho}H_{\alpha\beta\gamma}\partial_\sigma H_{\delta\epsilon\xi}\partial_\eta H_{\tau\phi\chi} $. This time they do correspond to genuinely new interactions and are not total derivatives, in difference to the case for the massless 1-form. However, for our case in $D=4$ dimensions this means that we cannot construct Galileon-type of interactions for the massless 2-form either. Hence, in four dimensions we can neither have Galileon interactions $f(F^2)\epsilon\epsilon(\partial F)^mF^n$ for the massless 1-form nor $f(H^2)\epsilon\epsilon(\partial H)^mH^n$ for the massless 2-form. 
  
Another way how we could construct interactions are via mixings between the massless 2-form and the massless spin-1 field. Thus, one could construct terms of the form $H^mF^n$. Since $H$ carries three but $F$ two indices, the first contribution starts at $m=2$ and $n=2$. This type of construction will nevertheless not generate Galileon interactions but only contribute to the quadratic Lagrangian $\mathcal{L}_2=f_2(B_{\mu\nu}, H_{\mu\nu\rho}, \bar{H}_{\mu})$ in the original formulation in terms of the massive 2-form. The attempt to construct derivative mixings like $\epsilon\epsilon H^m(\partial F)^n$ or $\epsilon\epsilon F^m(\partial H)^n$ face the same difficulty in four dimensions as the Galileon construction for the pure sectors, since one can construct them only starting from $D\ge6$ dimensions. Thus, it is not possible to construct Galileon-type of derivative self-interactions in the decoupling limit while keeping the gauge invariance for the 2- and 1-forms in four dimensions.
%%%%%%%%%%%%%%%%%%%%%%%%%
\section{First cosmological application}
\label{capp}
%%%%%%%%%%%%%%%%%%%%%%%%%

In this section we highlight some features of the $2-$form model and discuss the background evolution  in some particular simple models interesting for cosmology.  Previous interesting studies of $p-$forms with and without non-minimal couplings to gravity in the context of inflation and dark energy scenarios can be found in  \cite{Germani:2009iq,Koivisto:2009sd,Koivisto:2009fb,Koivisto:2012xm,Thorsrud:2012mu,Mulryne:2012ax,Ohashi:2013qba, Ohashi:2013mka,Ito:2015sxj,Kumar:2016tdn,Farakos:2017jme,Obata:2018ilf,Almeida:2019xzt,Almeida:2019iqp,Guarnizo:2019mwf,Takahashi:2019vax}.

%%%%%%%%%%%%%%%%%%%%%%%%%%%%%%%%%
\subsection{Gauge invariant}
The simplest example that we can consider is a theory based on a non-linear kinetic term. Since the kinetic term only depends on the gauge invariant field strength $H_{\alpha\beta\mu}$, the gauge symmetry will be intact. Such non-linear kinetic terms are very interesting since they are the simplest extension that one can consider and they also provide a K-mouflage screening mechanism together with a promising quantum behaviour. We  consider in this subsection a Lagrangian of a $2-$form minimally coupled to gravity: $\mathcal{L}_2=\sqrt{-g}\left[ \frac{M_{\rm pl}^2}{2}R + f_2(X)\right]$, where $X=-\frac1{12}H^2$. We use the FLRW Ansatz  
\begin{equation}
\mr{d} s^2=-N(t)^2\mr{d} t^2+a(t)^2\mr{d}\boldsymbol{x}^2\, , 
\end{equation}
for the metric and for the 2-form the background field configuration $B^{ij}=\frac13C\epsilon^{ijk}x_k$, where the remaining components all vanish. The associated energy density and pressure are given by 
\begin{equation}
\rho_B=-f_2 \qquad \text{and} \qquad p_B=f_2-2Xf_{2,X} \,.
\end{equation}
The Einstein's field equations are then simply given by 
\begin{equation}
3H^2=\rho_B \qquad \text{and} \qquad 2\dot{H}+3H^2=-p_B \,.
\end{equation}
The slow roll parameter $\epsilon=\frac32(w+1)=-\frac{\dot{H}}{H^2}$ indicates a regime with quasi-de Sitter solution when $\log f_{2,\log X}\ll1$. The covariant expression for the stress energy tensor $T_{\alpha\beta}=f_2g_{\alpha\beta}+\frac12H_{\alpha\rho\sigma}H_{\beta}{}^{\rho\sigma}f_{2,X}$ helps us to obtain quickly the propagation speed of scalar perturbations after introducing them in the metric and the 2-form. It is simply given by
\begin{equation}
c_s^2=1+2X\frac{f_{2,XX}}{f_{2,X}}\,.
\end{equation}
Similarly to the scalar counterpart of k-essence theories, the quasi-de Sitter regime would suffer from gradient instabilities.

\subsection{Non-minimal coupling}
In \cite{Heisenberg:2019akx} non-minimal couplings for the massive 2-form were investigated and it was shown that a unique coupling via the double dual Riemann tensor
arises
\begin{equation}\label{2formLBB}
\mathcal{L}^{\rm non-min}=\sqrt{-g}L^{\mu\nu\alpha\beta}B_{\mu\nu}B_{\alpha\beta}\,,
\end{equation}
where $L^{\mu\nu\alpha\beta}$ represents the double dual Riemann tensor $L^{\mu\nu\alpha\beta}=\frac14\epsilon^{\mu\nu\rho\sigma}\epsilon^{\alpha\beta\kappa\delta}R_{\rho\sigma\kappa\delta}$.
Additional support for this unique non-minimal coupling of the 2-form comes from the decoupling limit. After introducing the Stueckelberg field in \eqref{TrafoStuck} we have a massless 2-form and a massless spin-1 field, and the potential non-minimal couplings have to be valid couplings for these separate sectors. Since the massless spin-1 field has a unique non-minimally coupling to gravity via $L^{\alpha\beta\gamma\delta}F_{\alpha\beta}F_{\gamma\delta}$, this translates back to having \eqref{2formLBB} as the unique possible non-minimal coupling for the original massive 2-form.
Let us consider the following action
\begin{equation}\label{eq:act1}
\mathcal{S}=\int \mr{d}^4x \sqrt{-g}\left( \frac{M^2_{\rm{pl}}}{2}R-\frac1{12} H_{\mu\nu\rho} H^{\mu\nu\rho}-V[B_{\mu\nu}B^{\mu\nu}]+\gamma L^{\mu\nu\alpha\beta}B_{\mu\nu}B_{\alpha\beta} \right)\,,
\end{equation}
where $V[B_{\mu\nu}B^{\mu\nu}]$ is a general potential function for the $2-$form field. It is important to emphasize that the previous Lagrangian includes the topological $B\wedge F$ term (or, equivalently the term ${\cal L}_4$), through the  mechanism of topological generation of mass \cite{Allen:1990gb,Quevedo:1996uu,Dvali:2005ws} (see also \cite{Almeida:2018fwe, Almeida:2019xzt} for the discussion in curved background) reviewed in section \ref{sec:syspf}. \\
On top of a FLRW background the massive 2-form shall admit the following background field configuration
\begin{eqnarray}
B_{\mu\nu} =
\begin{pmatrix}
0&0&0&0 \\
0&0 &a^2 b&-a^2 b \\
0&-a^2 b&0 &a^2 b \\
0&a^2 b&-a^2 b&0
 \end{pmatrix}\,,
\end{eqnarray}
where the splitting of variables $a^2b=a(t)^2 b(t)$ is solely chosen for convenience.\\

The action in \cref{eq:act1} can then be brought up to total derivatives into the following symmetry reduced form
\begin{equation}\label{eq:act2}
\mathcal{S}=\int \mr{d}^4x \,\frac{3 a}{N} \left(
\frac{a^2b'^2}{2}
+2\Gamma\,a a' b b'
+\left(2\Gamma b^2-M_{\rm{pl}}^2\right)a'^2
 \right)
 - a^3 N V[6b^2]\,,
\end{equation}
where primes denote derivatives with respect to the time coordinate $t$ and we redefine the coupling constant $\gamma$ through $\Gamma \equiv 1 +4 \gamma$. 

Using the invariance of the reduced action under reparametrization of $t$, one can absorb the lapse function and rewrite the action and the equations of motion by defining the proper time $\tau$ as $\mr{d}\tau=N \mr{d}t$. Introducing the notation $\dot{a}=\mr{d}a/\mr{d}\tau$ and $\dot{b}=\mr{d}b/\mr{d}\tau$ the resulting background equations of motion are given by:
\begin{equation}
\begin{aligned}\label{eq:eombH}
\mathcal{E}_b=\; &\ddot{b}+3 H \dot{b}+2 b \left\{2 V'[6 b^2]+\Gamma
   \left(\dot{H}+H^2\right)\right\}=0 \,,\\
\mathcal{E}_N=\; &
H^2 \left(12 \Gamma b^2-6 M_{\rm{pl}}^2\right)+2
   V[6 b^2]+3 \dot{b}^2+12 \Gamma H b \dot{b} =0\,,\\
\mathcal{E}_a=\; &
\dot{H} \left(8 \Gamma b^2-4 M_{\rm{pl}}^2\right)
+H^2 \left(12 \Gamma b^2-6 M_{\rm{pl}}^2\right)
+2 V[6 b^2]
+4 \Gamma b\ddot{b}
+(4 \Gamma -3) \dot{b}^2
+16 \Gamma  H b \dot{b}=0\, .
\end{aligned}
\end{equation}
These can be brought into an autonomous form
\begin{equation}\label{auto}
\dot{H}=f_1 \qquad \text{and} \qquad \dot{b}=f_2 \; ,
\end{equation}
where we have defined:
\begin{equation}\label{autodef}
\begin{aligned}
f_1=\; &\frac{4 \Gamma\, H\,b \,f_2+(4 \Gamma -3) \,f_2^2-4 \Gamma\, b^2
   \left(4 V'+(2 \Gamma -3)\, H^2\right)+2 V-6
   H^2 M_{\rm{pl}}^2}{4 \left((2\Gamma-2)\Gamma\, b^2+M_{\rm{pl}}^2\right)} \, ,\\
   f_2=\; &-2 \Gamma\, H\, b \pm\sqrt{2 H^2 \left((2\Gamma-2)\Gamma b^2+M_{\rm{pl}}^2\right)-\frac{2}{3} V}\,.
\end{aligned}
\end{equation}
The critical points correspond to $\dot{H}=0$ and $\dot{b}=0$. For the stability analysis of the fixed points, we can consider small perturbations $\delta H$ and $\delta b$.
Defining the perturbation vector $v=\{\delta H, \delta b\}$ we can write the perturbations equations as $v'=\mathcal{M}v$, where the matrix $\mathcal{M}$ is given by
\begin{equation}\label{eq:Jacob}
\mathcal{M} =
\begin{pmatrix}
\frac{\partial f_1}{\partial H}&\frac{\partial f_1}{\partial b} \\[2mm]
\frac{\partial f_2}{\partial H}&\frac{\partial f_2}{\partial b}
 \end{pmatrix}\, , 
\end{equation}
evaluated at the critical point. In order for the critical point to be an attractor, all the eigenvalues $\lambda_i$ of the matrix $\mathcal{M}$ have to be negative, since the
perturbations in the environment of the critical point evolve as $e^{\lambda_i}t$.

\begin{figure}
\centering
\subfloat[\label{fig:SLP1}]{%
  \includegraphics[width=0.48\columnwidth]{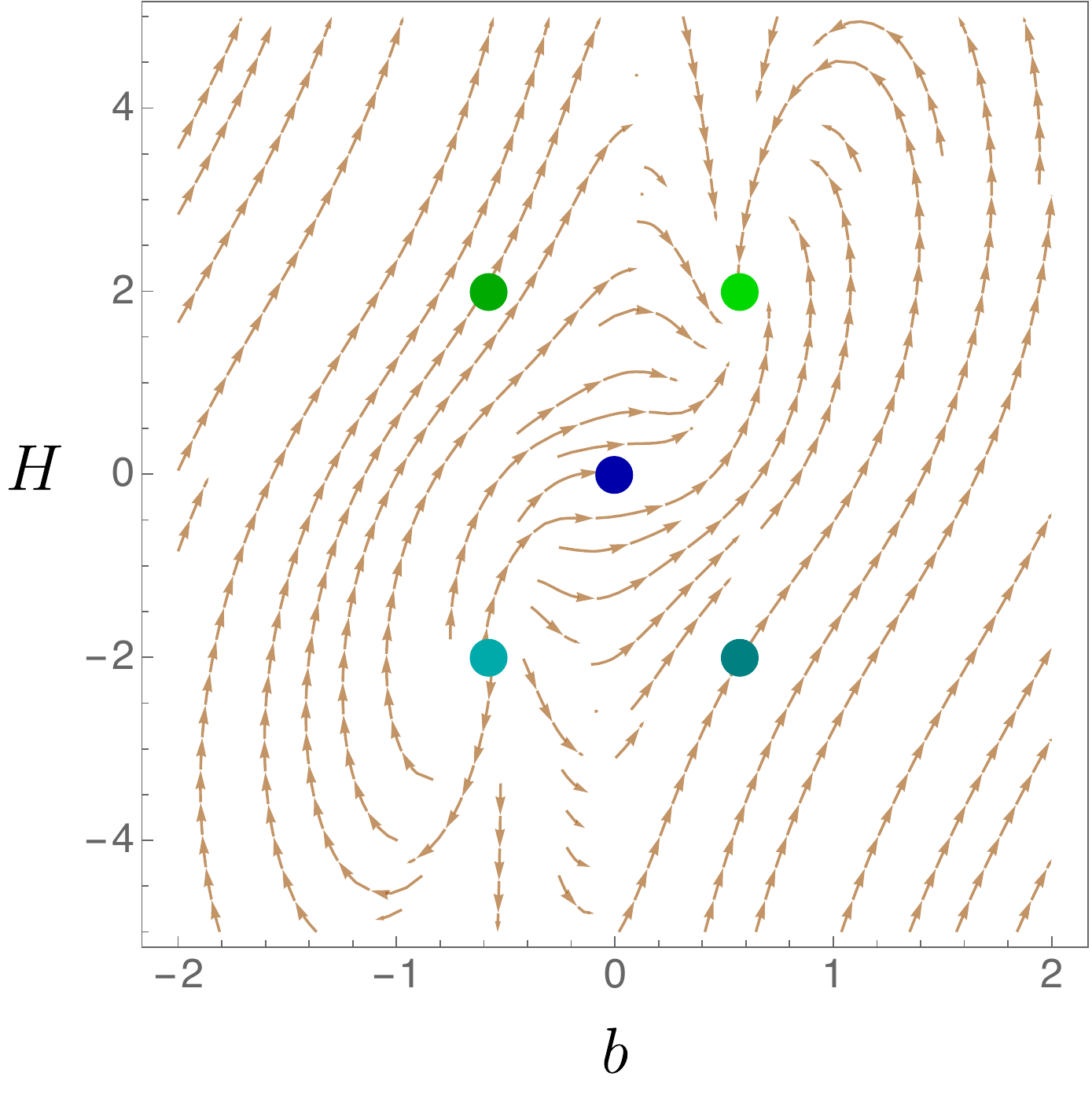}%
}\hfill
\subfloat[\label{fig:SLP2}]{%
  \includegraphics[width=0.48\columnwidth]{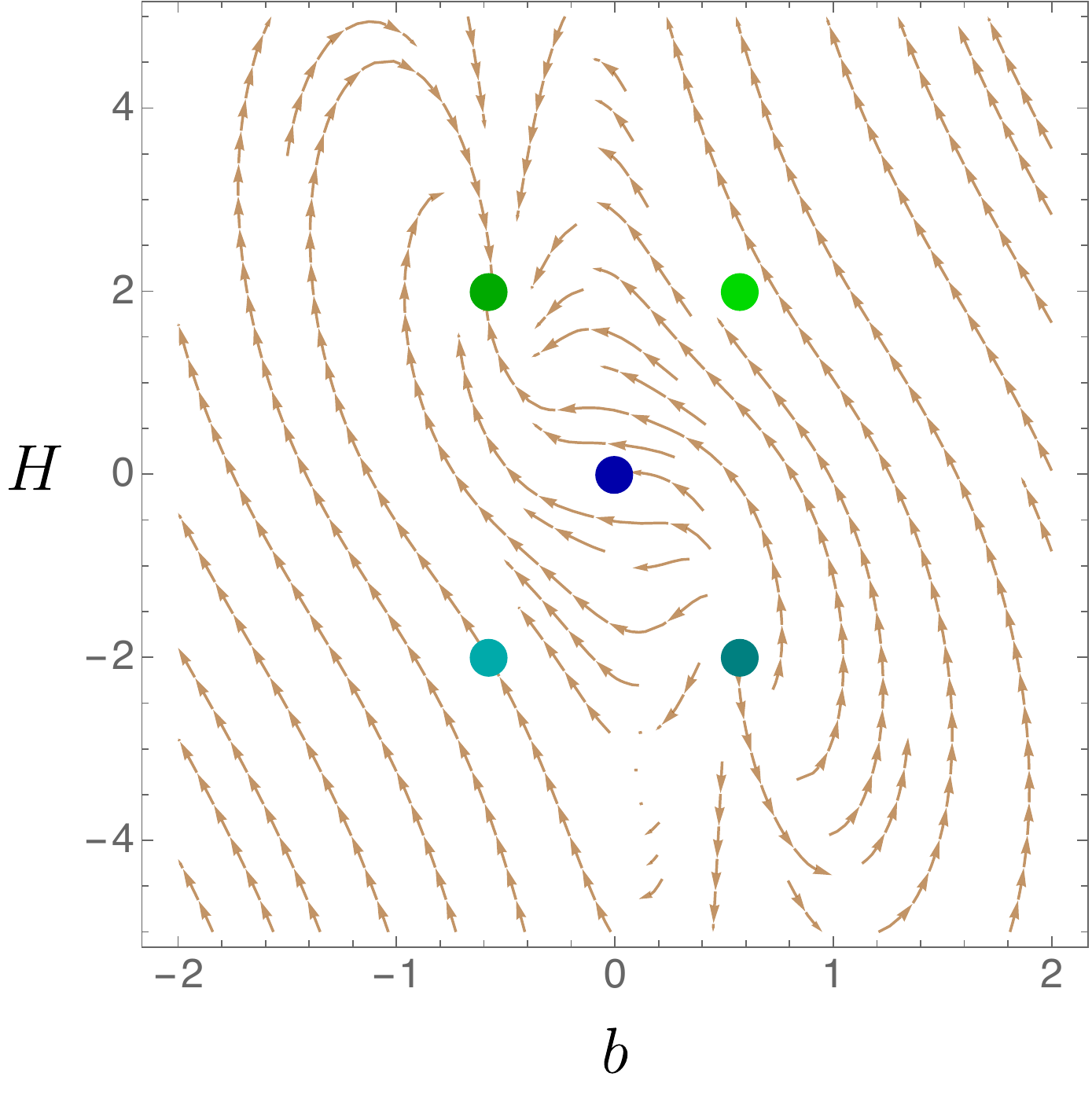}%
}
\caption{These plots show the dynamical phase portrait of the autonomous system of equations \eqref{auto} with an interacting potential term $V[x]=-g_b\,x^2$ for the case $g_b=1$, $\Gamma=2$ ($\gamma=\tfrac{1}{4}$) and $M_{\rm{pl}}=1$ in the phase plane $\{b,H\}$. The colored dots represent the five critical points \eqref{critP}. (a) The phase portrait of the $+$ sign choice in \eqref{autodef}. The positive critical point $\{b_c,H_c\}=\{\tfrac{1}{\sqrt{3}},2\}$ is a global attractor, such that the theory admits a stable critical de Sitter point. This means that the model as such is a successful dark energy theory candidate. Moreover, trajectories evolving from the repeller $\{b_c,H_c\}=\{-\tfrac{1}{\sqrt{3}},-2\}$, thus starting with $H<0$ which could model a contracting phase of the universe will all cross $H=0$ and end up in an expanding phase at the stable de Sitter attractor. This shows that the theory in principle as well represents a possible model of early universe scenarios without any initial singularity. (b) The phase portrait of the $-$ sign choice in \eqref{autodef} essentially shares the same characteristics in a mirrored manner.}
\label{fig:SLP}
\end{figure}

One can find non trivial critical points of the autonomous system \eqref{auto} by considering for example an interacting potential of the form $V[x]=-g_b\,x^2$.\footnote{Such that $V[6b^2]=-36g_b\,b^4$ and $V'[6 b^2]=-12g_b\,b^2$.} In this case, the system admits the five distinct critical points:
\begin{equation}\label{critP}
\{b_c,H_c\}=\{0,0\},\,\left\{\pm\frac{ \sqrt{\tfrac{2}{3}}\,M_{\rm{pl}}}{\sqrt{\Gamma}},\,\pm \frac{4\sqrt{g_b}\,M_{\rm{pl}}}{\Gamma}\right\},\,\left\{\mp\frac{ \sqrt{\tfrac{2}{3}}\,M_{\rm{pl}}}{\sqrt{\Gamma}},\,\pm \frac{4\sqrt{g_b}\,M_{\rm{pl}}}{\Gamma}\right\}\,,
\end{equation}
which are well defined as long as $\Gamma > 0$. Choosing for concreteness a unit coupling $g_b=1$ together with $\Gamma=2$ ($\gamma=\tfrac{1}{4}$) and using units in which $M_{\rm{pl}}=1$ one can analyze a concrete phase portrait of the dynamical system in the $\{b,H\}$ phase plane, depicted in Fig.\ref{fig:SLP1} for the $+$ sign choice in Fig.\eqref{autodef} and \ref{fig:SLP2} in the case of a $-$ sign with the five critical points represented as colored dots.

Focusing on the $+$ case, that is the left graph of Fig.\ref{fig:SLP}, one immediately observes that the positive critical point $\{b_c,H_c\}=\{\tfrac{1}{\sqrt{3}},2\}$ plays the role of a global attractor. This means that the theory admits a stable de Sitter solution which is essential for a possible application of the model to late time cosmology. The negative critical point, on the other hand, is a repeller, while the mixed ones together with the trivial fixed point are saddle points. In that sense, the phase portrait also indicates promising applications of the theory to singularity-free alternative early universe scenarios, in which trajectories starting at the global repeller $\{b_c,H_c\}=\{-\tfrac{1}{\sqrt{3}},-2\}$ initially possess a negative value of the Hubble parameter and thus describe a contracting phase of the universe. Without any exception, the trajectories will then at some point cross the value $H=0$ and inevitably end up in an expanding phase at the stable de Sitter critical point.

\begin{figure}
\centering
 \includegraphics[width=0.55\linewidth]{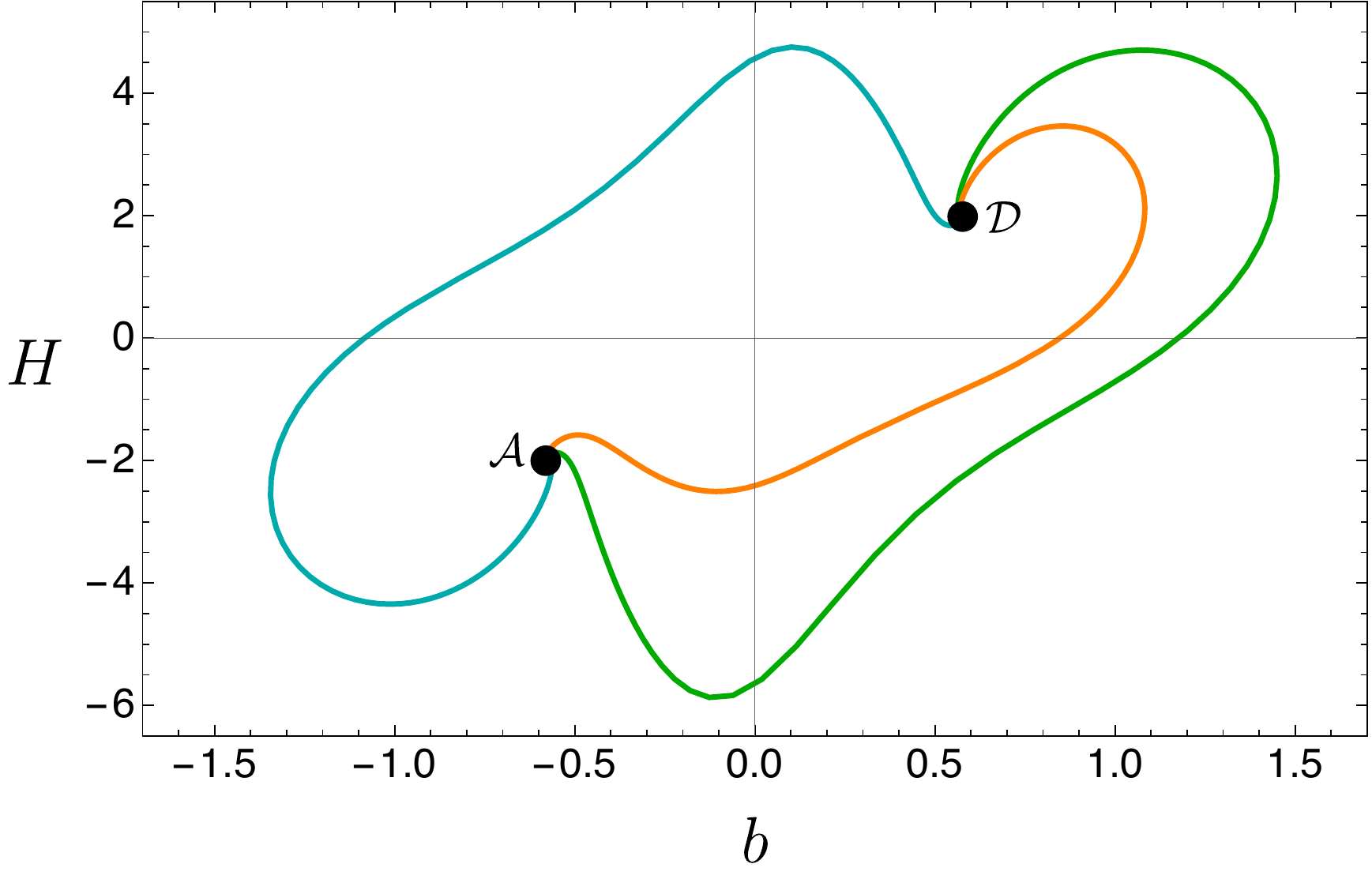}
\caption{This plot shows numerically integrated branches of solutions to the autonomous equations \eqref{auto} in the $+$ sign choice and choosing $g_b=1$, $\Gamma=2$ ($\gamma=\tfrac{1}{4}$) and $M_{\rm{pl}}=1$. The initial values of all three branches were chosen in the vicinity of the repeller $\{b_c,H_c\}=\{-\tfrac{1}{\sqrt{3}},-2\}$ ($\mL{A}$). All branches then evolve in distinct trajectories in the phase plane towards the stable de Sitter attractor at $\{b_c,H_c\}=\{\tfrac{1}{\sqrt{3}},2\}$ ($\mL{D}$). This represents a numerical check of the phase portrait consideration.}
\label{fig:PP}
\end{figure}

The above statements can readily be verified by solving the equations \eqref{auto} numerically. A parametric plot of three different solution branches which all start close to the global repeller ($\mL{A}$) is shown in Fig.\ref{fig:PP}. Every branch after very distinct trajectories winds up as expected at the global attractor ($\mL{D}$). A comparison with Fig.\ref{fig:SLP1} confirms the phase portrait results.

The $-$ sign case, that is \cref{fig:SLP2} shows qualitatively very similar results. The only difference is that the mixed critical points $\{b_c,H_c\}=\{\pm\tfrac{1}{\sqrt{3}},\mp2\}$ now play the role of the global attractor and repeller respectively.

\begin{figure}
\centering
\subfloat[\label{fig:WM1}]{%
  \includegraphics[width=0.48\columnwidth]{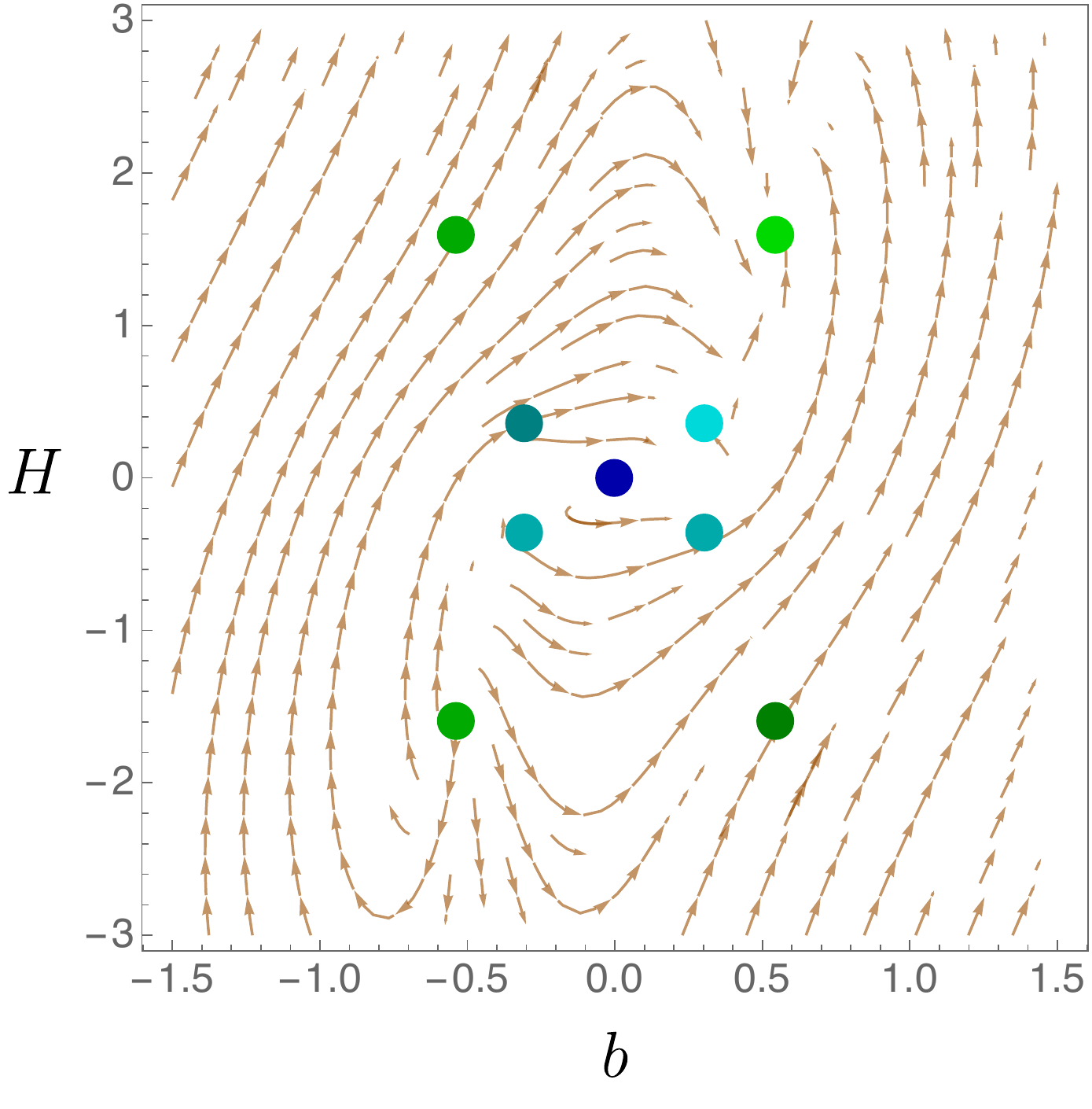}%
}\hfill
\subfloat[\label{fig:WM2}]{%
  \includegraphics[width=0.48\columnwidth]{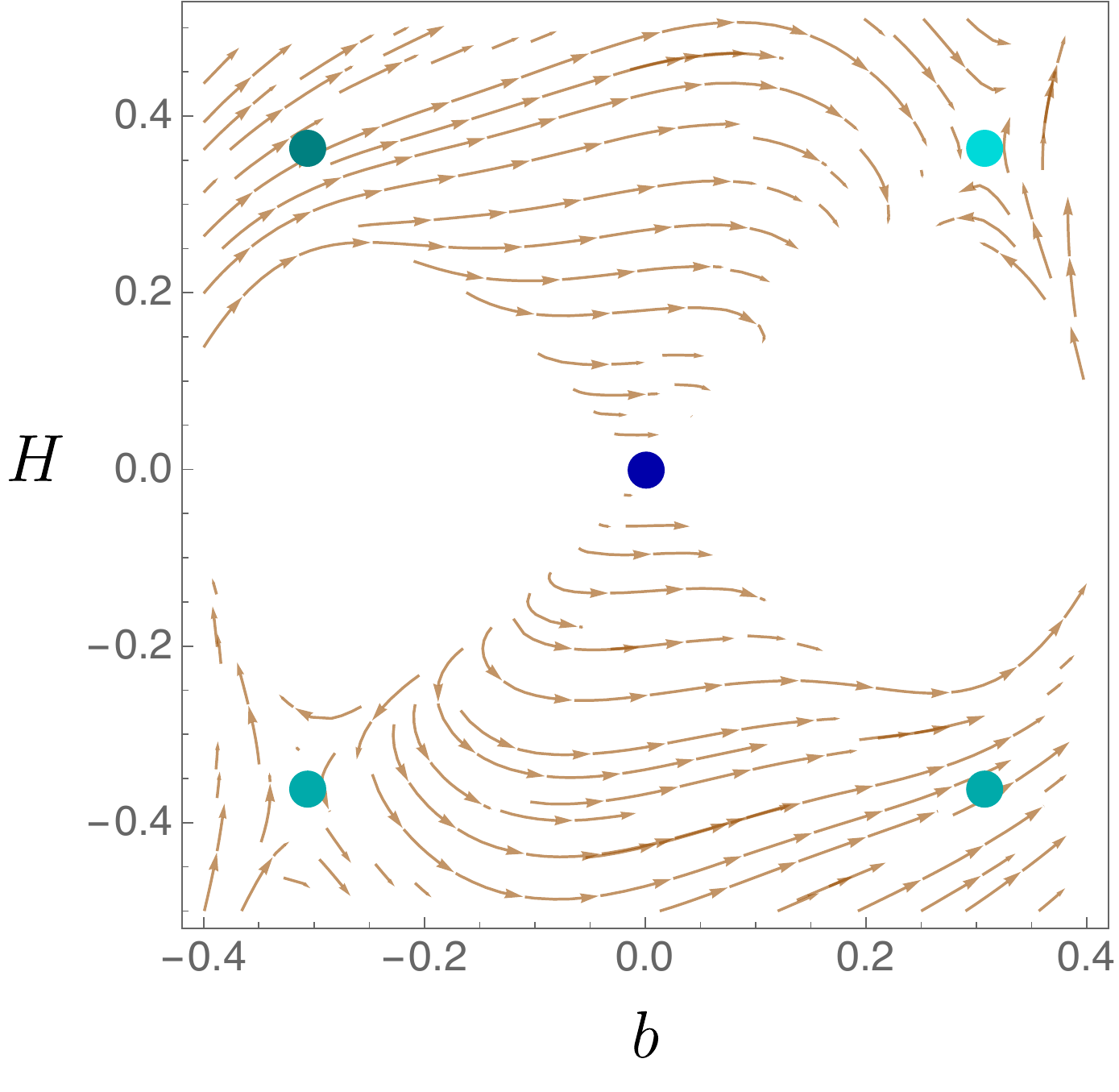}%
}
\caption{These plots show the dynamical phase portrait of the autonomous system of equations \eqref{auto} in the $+$ sign choice in \eqref{autodef} and an interacting potential with additional mass term $V[x]=\frac{1}{4}m^2\,x-g_b\,x^2$, with choices $m=2$, $g_b=1$, $\Gamma=2$ ($\gamma=\tfrac{1}{4}$) and $M_{\rm{pl}}=1$ in the phase plane $\{b,H\}$. The colored dots represent the nine critical points. This shows that additional terms in the choice of the potential have direct implications on the numbers of critical points and the shape of the solution space. (a) The overall picture far away from the central trivial critical point essentially remains the same and still admits stable as well as unstable critical de Sitter points. (b) An enlarged section of the phase portrait around the trivial critical point shows the direct effect of the additional mass term.}
\label{fig:WM}
\end{figure}

Choosing more involved potentials $V[6b^2]$ leads to even richer structures of solution space. For example, one can add a mass term $V[x]=\frac{1}{4}m^2\,x-g_b\,x^2$, such that $V[6b^2]=\frac{3}{2}m^2\,b^2-36g_b\,b^4$. This leads to additional four critical points which modify the phase portrait of the theory in a non trivial manner. As an illustrative example we will again choose specific coefficients $m=2$, $g_b=1$, $\Gamma=2$ ($\gamma=\tfrac{1}{4}$) and $M_{\rm{pl}}=1$. The corresponding phase protrait for a $+$ sign choice in the equation for the field $b$ is depicted in \cref{fig:WM}.

Far from the trivial critical point $\{b_c,H_c\}=\{0,0\}$, the picture remains qualitatively the same, with an attractor and a repeller as stable and unstable de Sitter solutions as shown in \cref{fig:WM1}. Near the null critical point, however, the picture changes significantly as seen in the enlarged phase portrait section in \cref{fig:WM2}. In particular, there are two spherical regions for which no real solution can be found.

Again a $-$ sign choice in \eqref{autodef} would lead to a mirrored picture of the above.

Of course these considerations should be viewed as preliminary checks for possible cosmological applications of the theory. We leave a more rigorous analysis including the matter sector for future work.

%%%%%%%%%%%%%%%%%%%%%%
\subsubsection{The conformal coupling case $\gamma=-1/4$}
%%%%%%%%%%%%%%%%%%%%%%
We end this section mentioning the special case of the coupling $\Gamma = 0 \; (\gamma = -1/4)$. From the system \eqref{eq:eombH} we see that this particular value leads to several simplifications. The simplified system reads: 
\begin{equation}\label{eq:eombH14}
\begin{aligned}
\mathcal{E}_b=\; &\ddot{b}+3 H \dot{b}+4  b   V'[6 b^2] =0,\\
\mathcal{E}_N=\; &
 -6 M_{\mr{pl}}^2 H^2 +2
   V[6 b^2]+3 \dot{b}^2 =0,\\
\mathcal{E}_a=\; &
 -4 M_{\mr{pl}}^2 \dot{H} 
-6 M_{\mr{pl}}^2 H^2  +2 V[6 b^2] -3 \dot{b}^2
 =0\, .
\end{aligned}
\end{equation}
It is worth to notice that for this value of $\Gamma$, the equation for the $2-$form is identical to the equation for a minimally coupled scalar field with a potential $V$ and identical to the vector inflation model  with non-minimal coupling of the form $RA_{\mu} A^{\mu}/6$ studied in \cite{Golovnev:2008cf}.  A closely related case of a $2-$form non-minimally coupled to gravity was also studied  in \cite{Germani:2009iq}. \\
IN order to see that such model can be relevant for the discussion of inflationary dynamics, we can compute the slow roll parameter $\epsilon$ for this system. Combine the second and the third equations we obtain
\begin{equation}\label{eq:srepsilon}
\epsilon = - \frac{\dot{H}}{H^2} = \frac{9 \dot{b}^2}{2V + 3\dot{b}^2} \approx \frac{3 \dot{b}^2}{2M_p^2 H^2 + \dot{b}^2},
\end{equation}
which tell us that for suitable potentials with $V \gg \dot{b}^2$ it is possible to sustain slow roll inflation. However, as extensively discussed in the literature \cite{Himmetoglu:2008zp,Himmetoglu:2009qi} this particular choice of coupling suffers from ghost instabilities in the longitudinal mode. A way out of instability problems relies on the inclusion of general kinetic couplings of the form $f(B^2) H_{\mu \nu\rho} H^{\mu \nu\rho}$ and general potential $V(B^2)$ as explored here. 

\section{Conclusion}
The construction of effective field theories is straightforward after determining the involved symmetries and the field content. In standard field theories, the usage of representations of the Lorentz group enables us to categorize the number of physical propagating degrees of freedom. A crucial difference arises between representations of massless and massive particles. A mass term generically breaks existing gauge symmetries of the massless limit and introduces additional propagating modes. A massless 1-form possesses, for instance, tow physical degrees of freedom, whereas its massive generalization introduces one additional degree of freedom due to broken $U(1)$ symmetry. There are different ways how these modes could be represented in alternative formulations. 

2-forms naturally arise in the low energy effective field theories of string theory. In this work, we investigated the topological mass generation of 2-forms and connected such a topological term to the recently proposed unique derivative coupling arising in the quartic Lagrangian of the systematic construction of massive $2-$form interactions. The massive 2-form finds a dual description in terms of a massless $1-$form and a massless $2-$form via a topological mass term $B\wedge F$. In this dual description the single degree of freedom propagated by the $2-$form is absorbed by the $1-$form, generating a massive mode for the $1-$form. There is a non-trivial correspondence between such a topological mass generation term and the massive 2-form interaction $\epsilon^{\mu\nu\rho\sigma}\epsilon^{\alpha\beta\gamma}_{\;\;\;\;\;\;\sigma}\partial_{\mu}B_{\alpha\rho}\partial_{\nu}B_{\beta\gamma}$ arising from the systematical construction in terms of the Levi-Civita tensor. This interaction is unique in the sense, that it represents a total derivative on its own but becomes a non-trivial interaction once an overall general function is introduced. Based on the decoupling limit analysis, we showed the uniqueness of such a topological mass term and absence of the Galileon-like interactions, in support of the arguments represented in \cite{Heisenberg:2019akx}. We also presented some preliminary applications in cosmology. 

\section*{Acknowledgments}
This work was partly supported by COLCIENCIAS -- DAAD grant 110278258747. JPBA and AG acknowledge support from Universidad Antonio Nari\~no grant  2019248. AG also aknowledges support from Universidad Antonio Nari\~no grant  2019101. LH is supported by funding from the European Research Council (ERC) under the European Unions Horizon 2020 research and innovation programme grant agreement No 801781 and by the Swiss National Science Foundation grant 179740. C.A.V.-T. acknowledges financial support from Vicerrector\'ia de Investigaciones (Univalle) grant 71220.
%%%%%%%%%%%%%%%%%%%%%%%%%%%%%%%%%%%%%%%%%%%%%%

\bibliographystyle{utcaps} 
\bibliography{bibli2form}

%\appendix

\end{document}